\documentclass[prl,preprint,graphicx,showpacs,preprintnumbers,amsmath,amssymb]{revtex4-1}

\usepackage{graphicx}
\usepackage{subfigure}
\usepackage{amsmath}

\draft 

\begin{document}


\title{A curved line search algorithm for atomic structure relaxation} 


\author{Zhanghui Chen}
\affiliation{Materials Sciences Division, Lawrence Berkeley National Laboratory, One Cyclotron Road, Mail Stop 50F, Berkeley, California 94720, United States}
\author{Jingbo Li}
\author{Shushen Li}
\affiliation{State Key Laboratory of Superlattices and Microstructures, Institute of Semiconductors, Chinese Academy of Sciences, P.O. Box 912, Beijing 100083, People's Republic of China}
\author{Linwang Wang}
\email{lwwang@lbl.gov}
\affiliation{Materials Sciences Division, Lawrence Berkeley National Laboratory, One Cyclotron Road, Mail Stop 50F, Berkeley, California 94720, United States}



\date{\today}

\begin{abstract}
 \emph{Ab initio} atomic relaxations often take large numbers of steps and long times to converge. An atomic relaxation method based on on-the-flight force learning and a corresponding new curved line minimization algorithm is presented to dramatically accelerate this process. Results for metal clusters demonstrate the significant speedup of this method compared with conventional conjugate-gradient method.
\end{abstract}


\maketitle 

One major usage of \emph{ab initio} density functional theory (DFT) in material science simulation is to determine the ground state atomic configuration for a given system \cite{NatPhy2009,RMP2000}. Overall, such applications probably take most of the DFT simulation time. There are two types of ground structure searching. The first is to find global minimum among many local minima \cite{Wales2001,clusters}. This has become an intensely studied topic in material design projects \cite{PRL2005,zhang2013,LBv,CALYPSO,USPEX}. Various types of evolutionary algorithms \cite{zhang2013,LBv,CALYPSO,USPEX,pdeco} or simulated annealing \cite{SA} schemes have been developed, as well as the minimum hopping methods \cite{PRL2005}. The second type is the conventional local minimum optimization, which is the concern of the current study. The related calculation is dominated by the conjugated gradient (CG) method \cite{CG,vasp,castep} and the Broyden-Fletcher-Goldfarb-Shanno (BFGS) method \cite{BFGS,abinit,pwscf}. Although these methods guarantee to converge into a local minimum, the convergence rate could be agonizingly slow, e.g. with hundreds of steps, thus a faster method will be extremely helpful. This local minimum problem also presents itself in the global minimum search since each global minimum search step usually deploys one or more local minimizations \cite{PRL2005,zhang2013,LBv,CALYPSO,USPEX}. One reason for the slow convergence of the local minimization steps is the possible narrow and curved energy valley leading to the minimum, which prevents the efficient execution of the conventional CG or BFGS methods. Imaging a rotation of a molecule on the surface of a substrate. Such rotation cannot be described by a straight line in cartesian coordinates which is used under CG or BFGS methods. In higher dimension, the situation can be more complicated, making it impossible to find the natural degree of freedom (e.g., the rotation angle). One such example is a metal cluster \cite{clusters,Cu20Au18,Co120} (which will be studied in this paper), where hundreds of steps might be needed to relax a structure while there is no obvious natural (or say internal) degree of freedom to speed up the convergence. To overcome these problems, one needs to do the minimization steps along guided curved lines following the energy valleys. We will call such algorithms the guided curved-line-search (CLS) algorithms.


The issue is how to find such guided curved lines. In this work, we will show that such guided curved line can be provided by model surrogate potentials with their parameters provided by on-the-flight fitting (OTFF) to the \emph{ab initio} atomic forces \cite{learn,forcematching,prb2010}. We will demonstrate the efficiency of our CLS algorithm on metal clusters. Overall, we have the following findings: (1) The CLS method can speed up the traditional CG method by a factor of 3 to 6 for both the number of steps and wall clock times; (2) The OTFF can be effectively used to speed up the atomic relaxations, not just molecular dynamics as it has been used so far \cite{learn,forcematching,prb2010};

As mentioned before, the guided curved line will be provided by a surrogate potential. One possible option is to carry out \emph{ab initio} line minimization along the steepest descent line (SDL) of this surrogate potential. We will use on-the-flight fitting (OTFF) to ensure that the atomic forces of this surrogate potential at the beginning of each step equal that of the \emph{ab initio} forces.  When the system approaching the final minimum point, the curved line will become straight in the small scale, then the curved line search will go back to the conventional straight line search. In practice, we found that the SDL can be warped with sharp twists in high dimensions. Besides, using SDL will miss the
 conjugated gradient feature between different line searches. To overcome these shortcomings, we will use the surrogate potential conjugate gradient descent line (SP-CGDL). To construct SP-CGDL, the conventional CG formalism is applied to the initial atomic force direction to yield the CG search direction. Then a straight line minimum search based on the surrogate potential is carried out. From the new line minimum point of the current surrogate function, subsequent CG straight lines are carried out. Thus, our SP-CGDL curved search line is consisted with many straight lines segments. The \emph{ab initio} line minimization will be carried out alone this SP-CGDL. One might worry that the \emph{ab initio} energy function along this segmented line might not be smooth enough to carry out \emph{ab initio} line minimization. But in practice, we found that one can effectively use the Brents algorithm \cite{brent} to search for the \emph{ab initio} line minimum along this SP-CGDL, and such line search often finds the line minimum at a few segments down the road along the SP-CGDL. Typically two \emph{ab initio} calculations are needed in the Brents algorithm to search for the minimum along the SP-CGDL \cite{LineSearch}, much like the conventional
line minimization calculation. After the \emph{ab initi}o line minimization are done, we call this one step, and the algorithm will repeat itself (from OTFF to construction of SP-CGDL, then \emph{ab initio} line minimization). Note, the above procedure maintains the feature of CG, when close to the minimum.

We will use metal cluster \cite{clusters,Cu20Au18,Co120} to demonstrate our CLS algorithm. The metal cluster potential is intrinsically high dimensional due to their long range atom-atom interaction. As a result, it is often difficult to reach their local minima. The metal cluster is an important subfield related to catalysts \cite{clusters,pdeco,Cu20Au18}. A lot of works have been done in searching of the optimal cluster structures, and the density of local minima in energy \cite{clusters,Cu20Au18,Co120,pdeco,PtKumar,PtXiao,Wang2009,PRL2005}. For metal systems, we found that the $N$-body Gupta force field \cite{Gupta} is a very good general potential. It has been used to model various types of metal clusters. The potential is a special case of the embedded atom potential \cite{EAM} based on the second moment approximation of the tight binding theory and it has the following form:
\[{E_N} = \sum\limits_{i = 1}^N {\left\{ {\sum\limits_{j = 1(j \ne i)}^N {{A_{ij}}\exp \left( { - {p_{ij}}\left( {\frac{{{r_{ij}}}}{{r_{ij}^0}} - 1} \right)} \right)}  - {{\left[ {\sum\limits_{j = 1(j \ne i)}^N {\xi _{ij}^2\exp \left( { - 2{q_{ij}}\left( {\frac{{{r_{ij}}}}{{r_{ij}^0}} - 1} \right)} \right)} } \right]}^{{1 \mathord{\left/
 {\vphantom {1 2}} \right.
 \kern-\nulldelimiterspace} 2}}}} \right\}} \]
where $r_{ij}$ represents the distance between the atom $i$ and $j$ in the cluster. The five groups of parameters $A_{ij}$, $\xi_{ij}$, $p_{ij}$, $q_{ij}$, $r_{ij}^0$ are allowed to vary independently to match the forces from DFT calculations. We restrict $A_{ij}=A_{ji}$, $\xi_{ij}=\xi_{ji}$, $p_{ij}=p_{ji}$, $q_{ij}=q_{ji}$ and $r_{ij}^0=r_{ji}^0$, and set them to zero when the $r_{ij}$ is larger than a cut-off distance to limit the number of the variables. Parameters are also restricted to vary within
a physically meaningful range.

Because the analytic expression for atomic forces of this model is a non-linear function of these parameters, in order to have an accurate force fitting, we have used a parallel differential evolutional algorithm \cite{pdeco} to globally minimize the force error. The resulting best solution is further optimized by a CG local minimization algorithm for these parameters. This approach enable us to
always fit the atomic forces with an error less than 0.005 eV/\AA, which is a few times lower than the typical \emph{ab initio}
minimization stoping criterion. Although the fitting procedure (at the beginning of every \emph{ab initio} line minimization step) might sound complicated, its computational cost is negligible, about 5$\%$ of the \emph{ab initio} computational time.

\begin{figure} [htbp]
\centering
\includegraphics[scale=0.7]{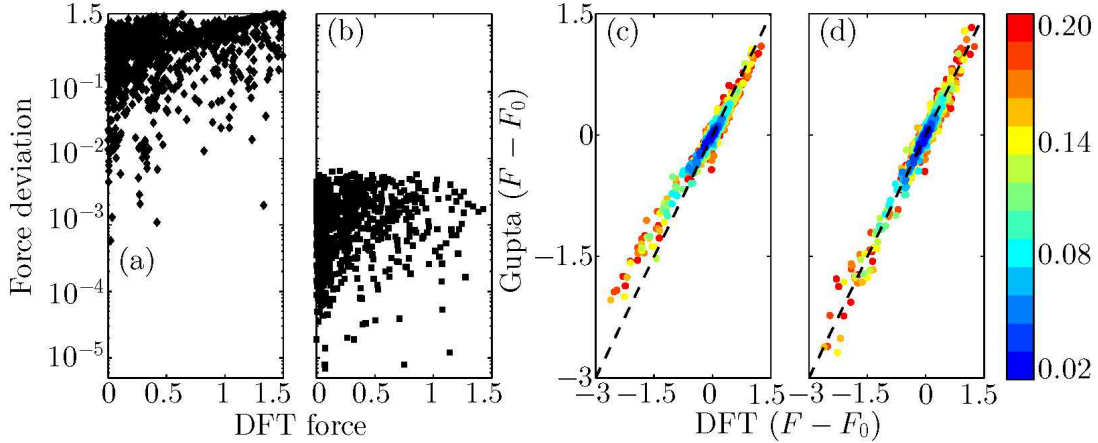}
\caption{Deviation of atomic forces for Pt$_{100}$ between DFT calculations and approximate models: (a) conventional un-fitted Gupta potential; (b) force-fitted Gupta potential. Comparisons of the values of force difference $F-F_0$ between DFT and approximate models: (c) un-fitted potential; (d) force-fitted potential, where $F_0$ is the force at the atomic structures ($R_0$) used for force fitting and $F$ is the one at randomly displaced structures ($R$) around $R_0$. Color bar for (c) and (d) indicates the distance (in unit of \AA) of $R$ from $R_0$. Note that the plot contains several different groups of $R_0$. In (a)-(b), the $x$-axis is DFT force while the $y$-axis is the deviation (in unit of eV/\AA). In (c) and (d), the $x$-axis is DFT force difference $F-F_0$ while the $y$-axis is the approximate one (in unit of eV/\AA).}\label{fig-force}
\end{figure}

To show the quality of atomic force fitting, we present the atomic force error in Fig.\ref{fig-force} for a Pt$_{100}$ cluster (with 100 Pt atoms). To begin with, we use the Gupta parameters from Ref.~\onlinecite{Gupta,Gupta2,GuptaParamaters} which have the parameters for almost all the major metallic elements. The atomic force error compared to \emph{ab initio} calculation using these original parameters without fitting is about 1 eV/\AA. After the parameter fitting, they becomes about 10$^{-3}$ eV/\AA. This improvement on the force is at no cost of degradation of other properties of this functional. For example, Fig.\ref{fig-force}(c),(d) compare the atomic force changes between the Gupta and DFT results when the atomic positions have been randomly displaced. The original Gupta result is already rather good, and it has been slightly improved after force fitting.

\begin{figure} [htbp]
\centering
\includegraphics[scale=0.65]{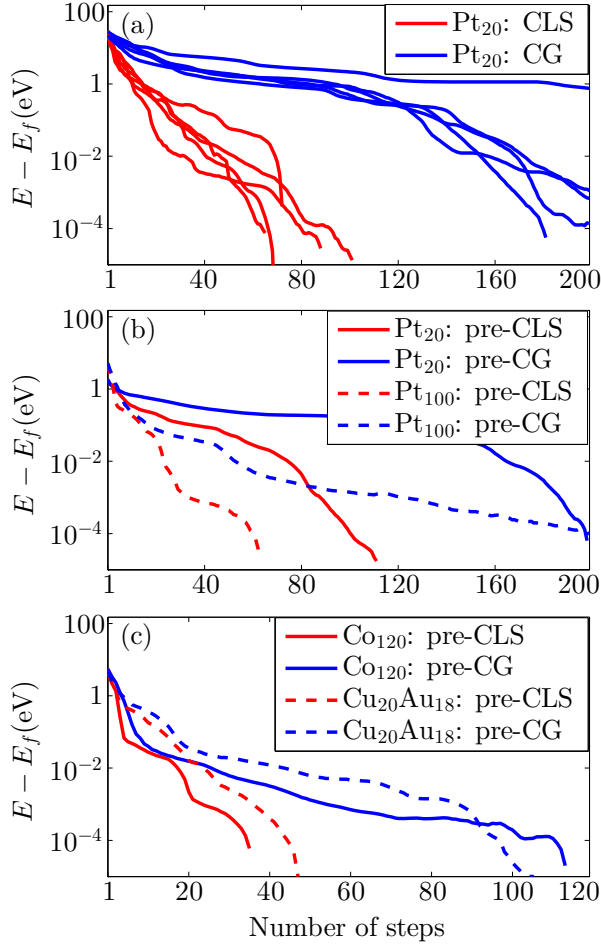}
\caption{(a) The relaxation process of five random Pt$_{20}$ clusters with different initial structures and corresponding different initial energies by the CLS and CG method, respectively; (b) the convergent process of the mean energy of these Pt$_{20}$ geometries after pre-relaxation, and the relaxation of Pt$_{100}$, (c) Co$_{120}$ and Cu$_{20}$Au$_{18}$ clusters with the initial geometry from the global minimum of conventional un-fitted Gupta force field. The $x$-axis is the number of relaxation steps while the $y$-axis is $E-Ef$ (in unit of eV), where $E$ is the energy of the current step and $E_f$ is the energy of the finally sought structure.}\label{fig-relax}
\end{figure}

To demonstrate the speedup of the CLS method, we first test five random Pt$_{20}$ clusters \cite{pdeco} with different initial structures and corresponding different initial energy. The convergence results are shown in Fig.\ref{fig-relax}(a) in comparison with the conventional CG results. We see that, more than a factor of 3 speedup are achieved for most cases. Especially, more than a factor of 6 speedup are achieved in the initial relaxation steps. Note, the total time for each line minimization step for CLS and the CG method is approximately the same.

In actual work, one often uses Gupta to pre-relax the system to a Gupta local minimum, then uses conventional \emph{ab initio} CG relaxation to further relax the total energy of the system. We will call such scheme pre-CG. One can also start with the Gupta relaxed minimum, then use our CLS method, we will call such method pre-CLS. Their results are shown in Fig.\ref{fig-relax}(b),(c). We can see that, due to the good approximation of Gupta to DFT energy, the Gupta minimum does provide a good initial speedup for \emph{ab initio} energy minimization, although pre-CLS still out performs pre-CG by a factor of 2 to 4, and if accurate results are needed, subsequent minimizations are as costly as the original CLS and CG methods. For the global minimization search problems, or to search for local minima density, one issue is that the pre-CG or pre-CLS method tends to mislead the system to the same local minimum near the Gupta potential basins for different initial configurations. This reduces the diversity of the global search. On the other hand, we found that the CLS method is exempted from such a problem.

Finally, in Fig.\ref{fig-relax}(c), we show that the CLS method also works for Co and CuAu alloy clusters, demonstrating its generality for metallic systems.

In summary, we have presented a curved line search (CLS) algorithm to speed up \emph{ab initio} atomic structure relaxation. This CLS uses a classical potential to provide the curved line on which\emph{ ab initio} line minimization is carried out. The parameters of this classical potential are fitted on-the-flight at every step to the \emph{ab initio} atomic forces. We tested this approach using metal clusters with Gupta force field as the classical potential and we expect similar approaches can be applied to other systems. Compared to the traditional CG method, we found CLS can speed up by a factor of 3-6.
\begin{acknowledgments}
The work of L.W. Wang is supported by the Material Theory program through the Director, Office of Science, Office of Basic Energy Sciences, Materials Science and Engineering Division, of the U.S. Department of Energy (DOE) under Contract No. DEAC02-05CH11231. This research used the resources of the National Energy Research Scientific Computing Center (NERSC) and Oak Ridge Leadership Computing Facility (OLCF) that are supported by the Office of Science of the U.S. Department of Energy, with the computational time allocated by the Innovative and Novel Computational Impact on Theory and Experiment (INCITEE) project. 
\end{acknowledgments}

%

\end{document}